\documentclass[lettersize,journal]{IEEEtran}

\usepackage{color}
\usepackage{epsfig}
\usepackage{cite}
\usepackage{subfigure}
\usepackage{multirow}
\usepackage{array}
\usepackage{mdwmath}
\usepackage{mdwtab}
\usepackage{amssymb,amsmath,amsthm}
\usepackage{enumerate}
\usepackage{graphicx}
\usepackage{setspace}
\usepackage{epstopdf}
\usepackage{comment}
\usepackage{texlogos}
\usepackage{cleveref}
\usepackage{amsmath}
\graphicspath{{Figures/}}
\usepackage{gensymb}
\usepackage{tablefootnote}
\usepackage{threeparttable}
\usepackage{mathtools}
\begin{document}
\title{A Robust Solver for Phasor-Domain Short-Circuit Analysis with Inverter-Based Resources}

\author{Aboutaleb~Haddadi,~\IEEEmembership{Senior~Member,~IEEE},~Evangelos~Farantatos,~\IEEEmembership{Senior~Member,~IEEE},~and~Ilhan~Kocar,~\IEEEmembership{Senior~Member,~IEEE}
%        and~Jane~Doe,~\IEEEmembership{Life~Fellow,~IEEE}% <-this % stops a space
\thanks{A. Haddadi and E. Farantatos are with the Electric Power Research Institute, Inc., 3420 Hillview Ave, Palo Alto, CA 94304, United States, e-mails: \{ahaddadi, efarantatos\}@epri.com.}% <-this % stops a space
\thanks{I. Kocar is with Polytechnique Montr\'{e}al, Montr\'{e}al, QC H3T1J4, Canada, e-mail: ilhan.kocar@polymtl.ca.}% <-this % stops a space
% \thanks{Manuscript received April 19, 2005; revised August 26, 2015.}
}

\maketitle

\begin{abstract}
The integration of Inverter-Based Resource (IBR) model into phasor-domain short circuit (SC) solvers challenges their numerical stability. To address the challenge, this paper proposes a solver that improves numerical stability by employing the Newton-Raphson iterative method. The solver can integrate the latest implementation of IBR SC model in industry-standard fault analysis programs including the voltage controlled current source tabular model as well as vendor-specific black-box and white-box equation-based models. The superior numerical stability of the proposed solver has been mathematically demonstrated, with identified convergence conditions. An algorithm for the implementation of the proposed solver in fault analysis programs has been developed. The objective is to improve the capability of the industry to accurately represent IBRs in SC studies and ensure system protection reliability in an IBR-dominated future.
\end{abstract}

\begin{IEEEkeywords}
Convergence of numerical methods, Inverter-based resource, Nonlinear network analysis, Numerical analysis, Numerical stability, Power system protection, Short circuit currents.
\end{IEEEkeywords}

\IEEEpeerreviewmaketitle

\section{Introduction}

\IEEEPARstart{T}{he} increased uptake of Inverter-Based Resources (IBRs) in the power system has precipitated a growing necessity for accurate modeling of these resources across various time-scale studies including short-circuit (SC)~\cite{Ref:Aidan100, Ref:PSRCWTG, Ref:C24Conf, Ref:HaddadiLoad}. Despite recent advancements~\cite{Ref:C24, Ref:Type4Model, Ref:EPRI_SC_2023, Ref:Type3Model, Ref:BESSmodel, Ref:HaddadiVCCS, Ref:EPRI_equiv}, integrating IBR SC models into a phasor-domain SC solver remains challenging~\cite{Ref:ManishNumerical, Ref:ManishTPWRD}. A primary challenge is maintaining the numerical stability of the solver~\cite{Ref:ManishNumerical,Ref:C24,Ref:ManishTPWRD}. The inherent nonlinearity of IBR fault ride-through (FRT) control and current limiter schemes~\cite{Ref:Hooshyar_FRT2, Ref:Hooshyar_FRT3, Ref:Hooshyar_FRT4, Ref:Maxime_FRT} imparts a nonlinear characteristic to the IBR SC model, necessitating iterations with the network solver. Under high IBR levels, this nonlinearity may reduce the numerical stability of the iterative mechanism, potentially leading to non-convergence~\cite{Ref:ManishNumerical,Ref:C24,Ref:ManishTPWRD}.

The literature has studied the numerical stability challenges of traditional SC solvers caused by integrating IBR models. Reference~\cite{Ref:ManishNumerical} has presented scenarios where the iterative method for calculating fault current contribution from an IBR SC model fails to converge. Reference~\cite{Ref:ManishTPWRD} has explored the numerical challenges of integrating IBR models into a SC model developed for systems dominated by synchronous machines. Reference~\cite{Ref:C24} has presented methods to improve the convergence of a traditional SC solver under IBRs. These studies highlight the need for specialized nonlinear network analysis methods tailored to IBR characteristics to maintain the solution integrity of traditional SC solvers. 

This paper proposes a Newton-Raphson (NR)-based phasor-domain solver for enhancing the numerical stability of a SC solver under IBRs. Initially, the paper identifies the poor numerical stability of the traditional IBR SC model~\cite{Ref:C24} as the root cause of numerical stability issues. Subsequently, a modified IBR SC model has been proposed, offering superior numerical stability compared to the traditional model. To integrate the proposed model, two variations of the proposed solver, denoted as Solver 1 and Solver 2, have been developed, each tailored to different types of IBR SC models. Solver 1, which assumes the availability of a voltage-controlled current source (VCCS) tabular model~\cite{Ref:C24}, implements the iterative NR method~\cite{Ref:BookStoer}. Solver 2, which is agnostic to the type of IBR model, implements a secant-based variation of the NR method~\cite{Ref:BookStoer}. A mathematical proof of convergence of the proposed solvers has been provided, and convergence conditions have been identified. Finally, an algorithm has been developed that could be used to implement the proposed solvers in fault analysis programs. The objective is to enhance the ability of the industry to accurately represent IBRs in SC studies, identify their potential impacts on system protection, and ensure the reliability of system protection as the power system transitions towards an IBR-dominated future.

\section{Numerical Stability Challenges of A Traditional Short Circuit Solver under Inverter-Based Resources}
\label{sec:Trad_solver}

\begin{figure}[!t]
\centering
\includegraphics[width=0.32\textwidth,trim={0 0 0 0},clip]{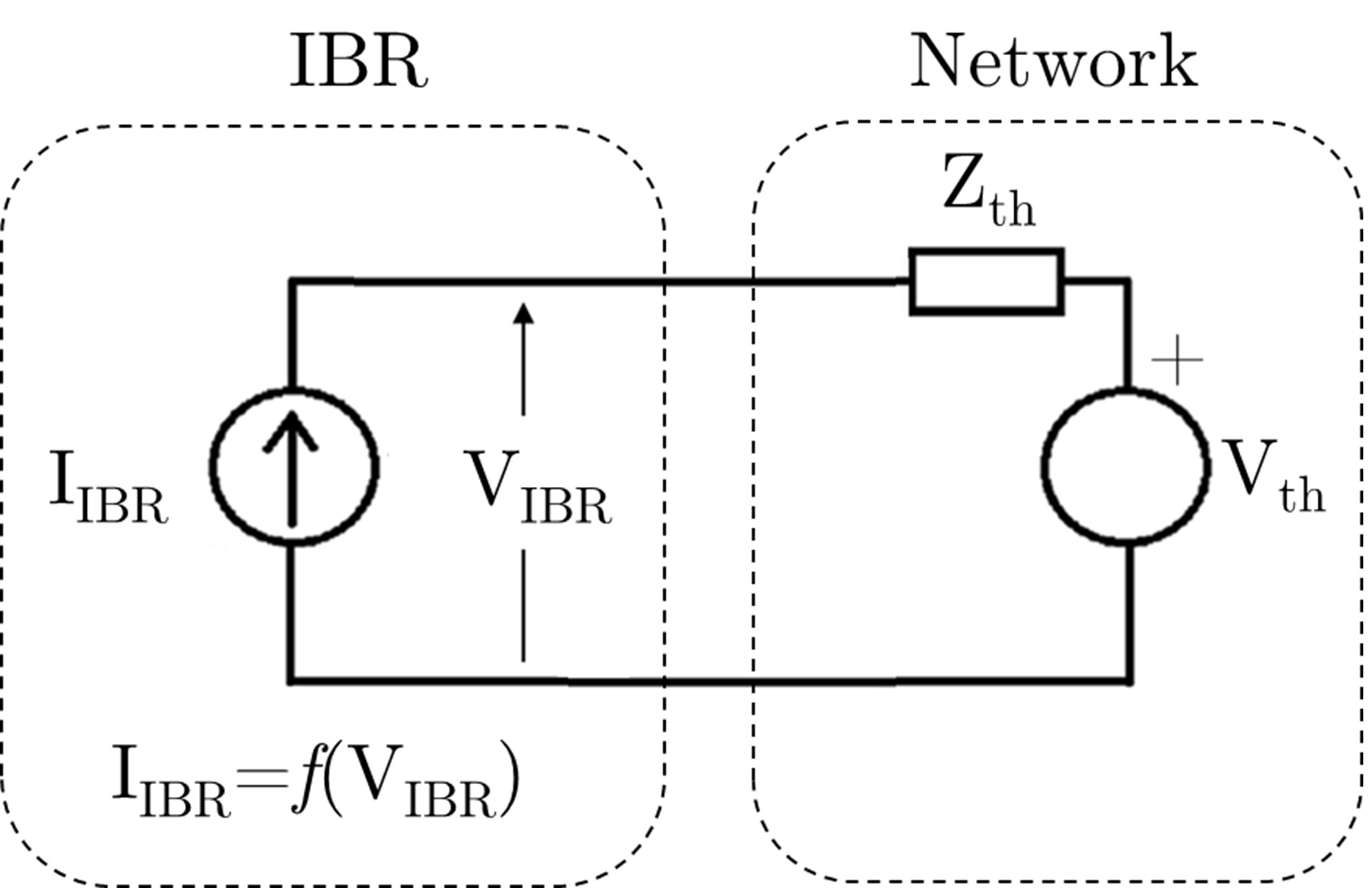}
\caption{Equivalent circuit employed by a traditional phasor-domain SC solver to determine the fault current contribution of an IBR model.}
\label{fig:Eq_Net_Trad_Sol}
\end{figure}

Figure~\ref{fig:Eq_Net_Trad_Sol} illustrates the equivalent circuit employed by a traditional phasor-domain SC solver to determine the fault current contribution of an IBR model~\cite{Ref:C24}. The IBR has been represented by an ideal current source~\cite{Ref:C24}, where the amplitude and phase angle of the output current complex phasor, $\textrm{I}_\textrm{IBR}$, depend on the terminal voltage complex phasor, $\textrm{V}_\textrm{IBR}$. This relationship can be expressed as
\begin{align}
\label{eq:IBR_control_eq}
\textrm{I}_\textrm{IBR} &= f(\textrm{V}_\textrm{IBR})\textrm{,}
\end{align}
where function $f(.)$ encompasses factors influencing the fault current of an IBR, including FRT control and current limiter schemes. Due to the nonlinearity of these schemes, $f$ is typically a nonlinear function. The network, as observed from the IBR terminal, has been modeled using a Thevenin equivalent in phasor domain, with $\textrm{V}_\textrm{th}$ and $\textrm{Z}_\textrm{th}$ representing the complex phasor of Thevenin voltage and Thevenin impedance, respectively. The Kirchhoff's voltage law (KVL) equation for the network is given by
\begin{align}\label{eq:network_KVL}
\textrm{V}_\textrm{IBR} &= \textrm{Z}_\textrm{th} \cdot \textrm{I}_\textrm{IBR} + \textrm{V}_\textrm{th}\textrm{.}
\end{align}
Calculating the fault current contribution of the IBR involves solving (\ref{eq:IBR_control_eq}) and (\ref{eq:network_KVL}). Given that (\ref{eq:IBR_control_eq}) is generally nonlinear, SC solvers employ an iterative method to solve these equations successively.

\begin{figure}[!t]
\centering
\includegraphics[width=0.4\textwidth]{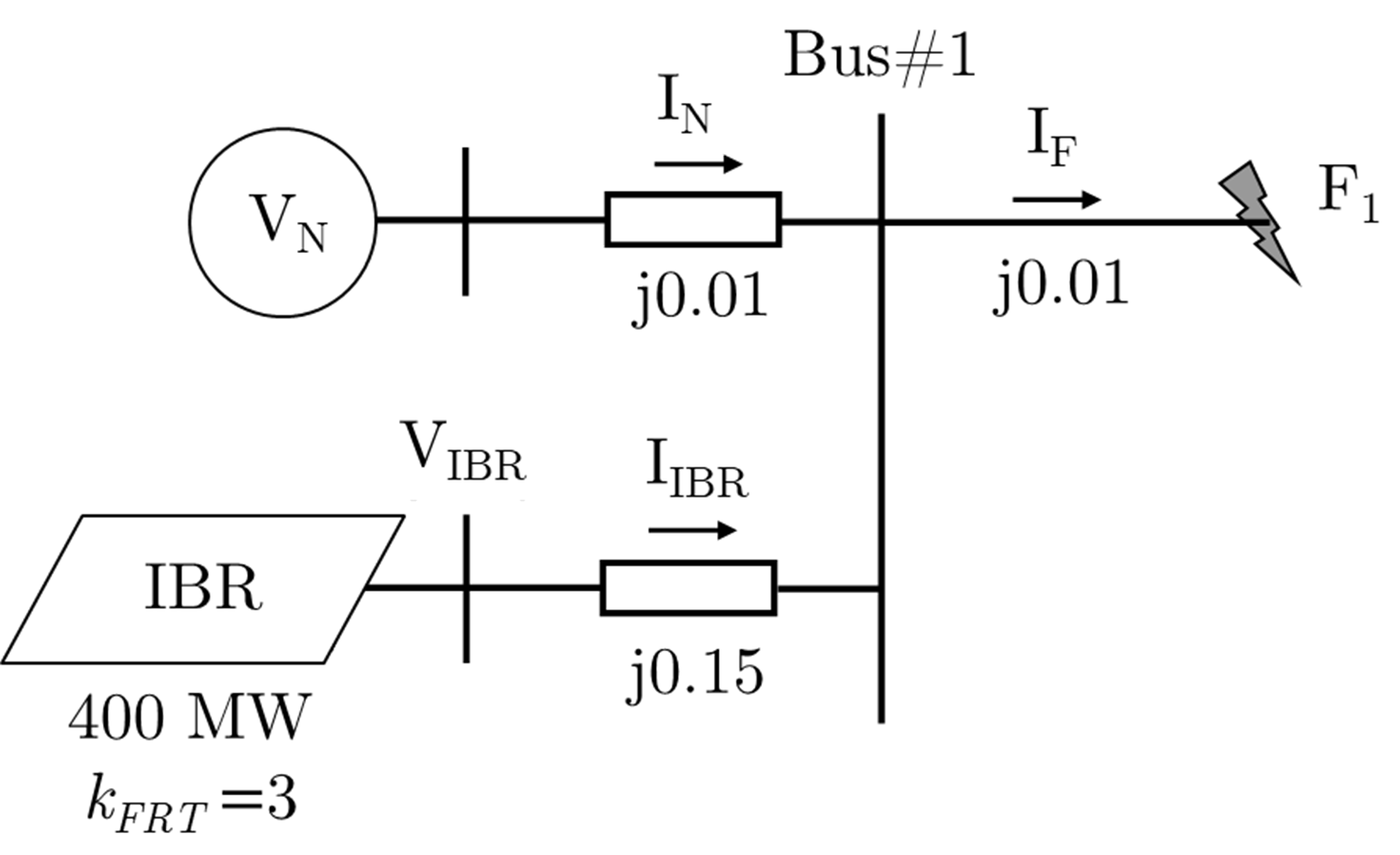}
\caption{Simple test system (the parameters are in per unit (pu) at 100 MVA base.)}
\label{fig:simpleTestSystem}
\end{figure}

\begin{table}[!t]\footnotesize
\renewcommand{\arraystretch}{1.3}
\caption{The VCCS Tabular Model of IBR in Fig.~\ref{fig:simpleTestSystem} at 400 MVA Base.}
\label{table:VCCS}
\centering
\begin{tabular}{m{8em}|m{8em}|m{9em}}
\hline
\textbf{Positive-sequence $\textrm{V}_\textrm{IBR}$ (pu)} & \textbf{Positive-sequence $\textrm{I}_\textrm{IBR}$ (pu)} & \textbf{Angle($\textrm{I}_\textrm{IBR}$/$\textrm{V}_\textrm{IBR}$) (Degrees)} \\
\hline
1.00 &	0.90 &	0.00\\
\hline
0.90 &	1.04 &	-16.70\\
\hline
0.80 &	1.20 &	-30.00\\
\hline
0.70 &	1.20 &	-48.59\\
\hline
0.60 &	1.20 &	-90.00\\
\hline
0.50 &	1.20 &	-90.00\\
\hline
0.40 &	1.20 &	-90.00\\
\hline
0.30 &	1.20 &	-90.00\\
\hline
0.20 &	1.20 &	-90.00\\
\hline
0.10 &	1.20 &	-90.00\\
\hline
\end{tabular}
\end{table}

To illustrate potential numerical challenges, the test system of Fig.~\ref{fig:simpleTestSystem} has been solved using the iterative method of a traditional SC solver. The test system represents a portion of a 230-kV transmission system including a 400 MW solar photovoltaic (PV)-based park connected to bus Bus\#1. The FRT control of the IBR operates in reactive current priority mode and provides dynamic reactive current control based on a k-factor control with $k_{FRT}=\textrm{3}$. References~\cite{Ref:WesPES23,Ref:PVMOD} have detailed the IBR control scheme. In phasor domain, the IBR has been represented by its VCCS tabular model of Table~\ref{table:VCCS}. The rest of the transmission system has been represented by a voltage source with a voltage of $\textrm{V}_\textrm{N}=1\angle 0\degree$ pu behind an impedance. A bolted three-phase-to-ground fault denoted by $\textrm{F}_\textrm{1}$ has been applied near bus Bus\#1. 

The considered traditional SC solver iteratively addresses  (\ref{eq:IBR_control_eq}) and (\ref{eq:network_KVL}); in each iteration, the solution of $\textrm{V}_\textrm{IBR}$ from the previous iteration is used to update $\textrm{I}_\textrm{IBR}$ in (\ref{eq:IBR_control_eq}). The updated IBR current injection is then substituted back into (\ref{eq:network_KVL}) to find the new $\textrm{V}_\textrm{IBR}$. This process is repeated until a stopping criterion is met. 

Table~\ref{table:iteration_trad} presents the first 20 iterations of this traditional solver. As shown, the iteration does not converge due to sustained numerical oscillation. The amplitude of $\textrm{V}_\textrm{IBR}$ oscillates between 0.41 pu and 0.98 pu, and the power factor angle of $\textrm{I}_\textrm{IBR}$ oscillates between -90\degree and zero; when the amplitude of $\textrm{V}_\textrm{IBR}$ drops to 0.41 pu, the IBR control injects a purely reactive $\textrm{I}_\textrm{IBR}$ to boost the voltage. However, this purely reactive current leads to a large increase in terminal voltage, resulting in $|\textrm{V}_\textrm{IBR}|=\textrm{0.98}$ pu. To correct this large voltage increase, in the next iteration the IBR adjusts the power factor angle of $\textrm{I}_\textrm{IBR}$ to zero, thus injecting a predominantly active current. This aggressive reduction in the injected reactive current, in turn, leads to a large drop in voltage, resulting in $|\textrm{V}_\textrm{IBR}|=\textrm{0.41}$ pu. This numerical oscillation continues until the maximum number of iterations is reached, at which point the solver declares non-convergence.

\begin{table}[!t]\footnotesize
\renewcommand{\arraystretch}{1.3}
\caption{Iterations of the Traditional Short-Circuit Solver for the Test System of Fig.~\ref{fig:simpleTestSystem}.}
\label{table:iteration_trad}
\centering
\begin{tabular}{m{6em}|m{8em}|m{8em}}
\hline
Iteration {\#} & $\textrm{V}_\textrm{IBR}$ (pu) &	$\textrm{I}_\textrm{IBR}$ (pu)\\
\hline
0 &	0.50$\angle$0.00\degree & 0 \\
\hline
1 &	1.27$\angle$0.00\degree & 1.20$\angle-$90.00\degree \\
\hline
2 &	0.77$\angle$49.22\degree & 0.90$\angle$0.00\degree \\
\hline
3 &	0.82$\angle$66.43\degree & 1.20$\angle$12.80\degree \\
\hline
4 &	0.58$\angle$87.61\degree & 1.16$\angle$39.38\degree \\
\hline
5 &	0.94$\angle$55.42\degree & 1.20$\angle-$2.39\degree \\
\hline
6 &	0.45$\angle$83.62\degree & 0.99$\angle$45.03\degree \\
\hline
7 &	0.97$\angle$52.66\degree & 1.20$\angle-$6.38\degree \\
\hline
8 & 0.42$\angle$82.63\degree & 0.95$\angle$46.98\degree\\
\hline
9 &	0.97$\angle$51.99\degree & 1.20$\angle-$7.37\degree \\
\hline
10 & 0.41$\angle$82.33\degree & 0.94$\angle$47.44\degree \\
\hline
11 & 0.97$\angle$51.78\degree & 1.20$\angle-$7.67\degree \\
\hline
12 & 0.41$\angle$82.24\degree & 0.94$\angle$47.58\degree \\
\hline
13 & 0.98$\angle$51.72\degree & 1.20$\angle-$7.76\degree \\
\hline
14 & 0.41$\angle$82.21\degree & 0.94$\angle$47.62\degree \\
\hline
15 & 0.98$\angle$51.69\degree & 1.20$\angle-$7.79\degree \\
\hline
16 & 0.41$\angle$82.20\degree & 0.94$\angle$47.64\degree \\
\hline
17 & 0.98$\angle$51.69\degree & 1.20$\angle-$7.80\degree \\
\hline
18 & 0.41$\angle$82.19\degree & 0.94$\angle$47.64\degree \\
\hline
19 & 0.98$\angle$51.69\degree & 1.20$\angle-$7.81\degree \\
\hline
20 & 0.41$\angle$82.19\degree & 0.94$\angle$47.65\degree \\
\hline
\end{tabular}
\end{table}

The numerical oscillation observed in the aforementioned example is caused by the iterative method employed by the traditional SC solver, rather than being indicative of an actual power system instability. To illustrate this, the test system of Fig.~\ref{fig:simpleTestSystem} has been simulated in time domain using an electromagnetic transient-type (EMT) solver employing a properly parameterized generic EMT model of the IBR~\cite{Ref:PVMOD}. The results, presented in Table~\ref{table:solution_simplesystem}, suggest the existence of a feasible phasor-domain solution. However, the iterative method of the traditional SC solver fails to converge to this solution.

\begin{table}[!t]\footnotesize
\renewcommand{\arraystretch}{1.3}
\caption{The Phasor-Domain Solution of the Test System of Fig.~\ref{fig:simpleTestSystem} at 400 MVA Base Obtained from an EMT Simulation.}
\label{table:solution_simplesystem}
\centering
\begin{tabular}{m{4em}|m{11em}}
\hline
\textbf{Quantity} & \textbf{Solution (pu)}\\
\hline
$\textrm{V}_\textrm{IBR}$ & 0.71 $\angle$73.48\degree \\
\hline
$\textrm{I}_\textrm{IBR}$ & 1.20 $\angle$23.50\degree \\
\hline
$\textrm{V}_\textrm{N}$	& 1.00 $\angle$0.00\degree \\
\hline
$\textrm{I}_\textrm{N}$	& 12.75 $\angle-$92.69\degree \\
\hline
$\textrm{V}_\textrm{Bus\#1}$	& 0.49 $\angle$2.36\degree \\
\hline
$\textrm{I}_\textrm{F}$	& 12.27 $\angle-$87.64\degree \\
\hline
\end{tabular}
\end{table}

In summary, the iterative method of SC solvers may fail to converge when applied to a traditional IBR SC model. This issue arises due to fluctuations and abrupt changes in the output current and terminal voltage of the traditional IBR model across successive iterations. These numerical oscillations are primarily due to the use of an ideal current source in the traditional IBR model.

\section{The Proposed Solver}

\begin{figure}[!t]
\centering
\includegraphics[width=0.4\textwidth]{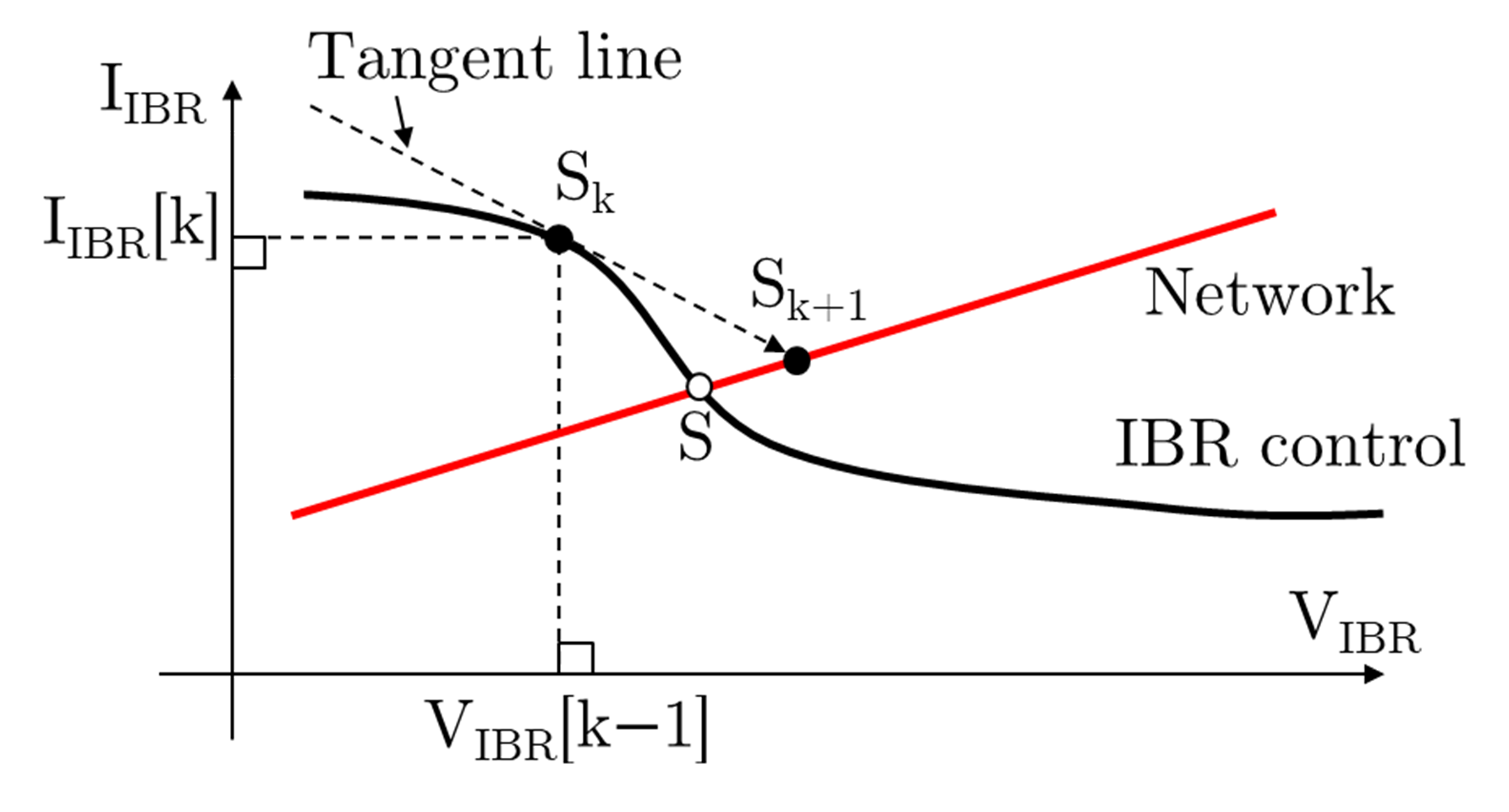}
\caption{Geometrical interpretation of the iteration of the proposed solver.}
\label{fig:Geometric_NR}
\end{figure}

\begin{figure}[!t]
\centering
\includegraphics[width=0.26\textwidth, trim={0 0 0 0},clip]{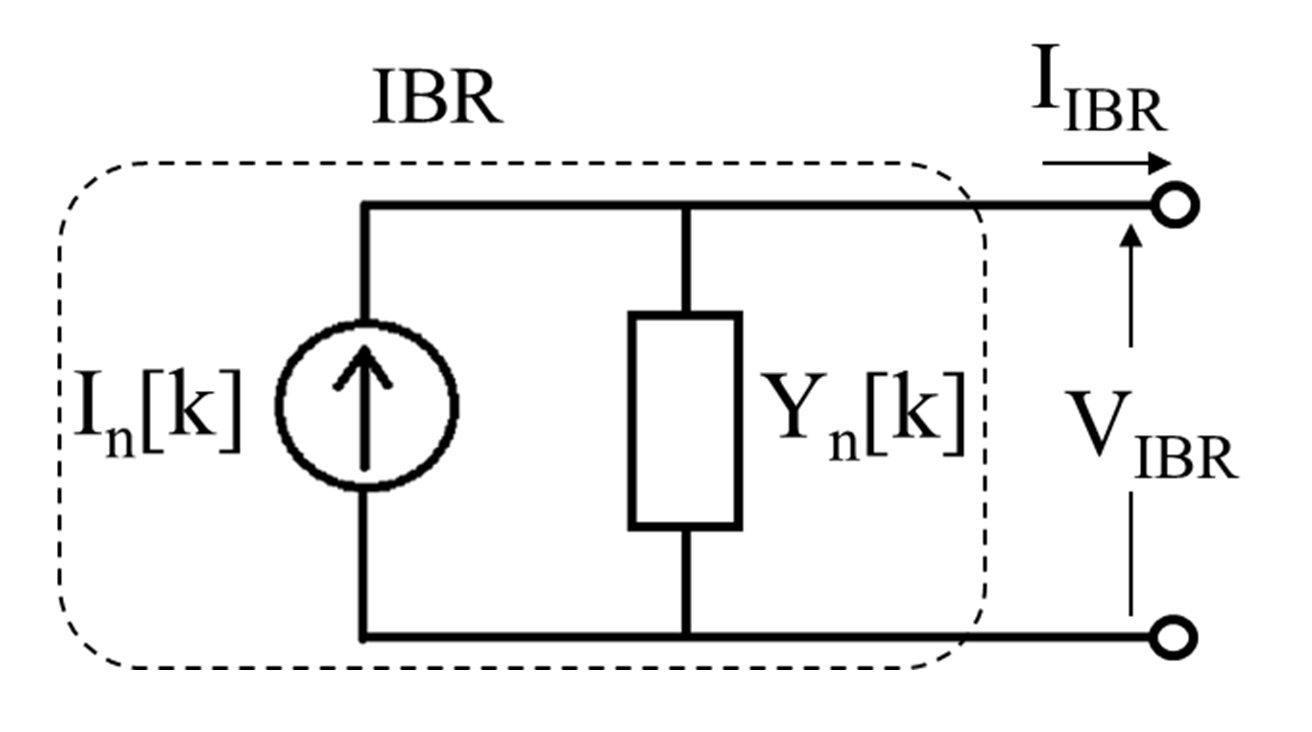}
\caption{The proposed IBR SC model.}
\label{fig:Modified_IBR_Model}
\end{figure}

To address the numerical stability challenge, this section develops a SC solver implementing the NR iterative method. Figure~\ref{fig:Geometric_NR} provides a conceptual geometrical interpretation of the iteration of the proposed solver. The black curve corresponds to the IBR control equation per (\ref{eq:IBR_control_eq}), while the red curve corresponds to the network equation per (\ref{eq:network_KVL}). If a solution exists, the two curves intersect at least once, marked by point S. In each iteration, the proposed solver defines the solution of the network equation by drawing a tangent line on the graph of the IBR control curve at the current solution point $\textrm{S}_\textrm{k}$ and moving to the next solution point $\textrm{S}_\textrm{k+1}$ on the network curve along this tangent line. In a given iteration $\#\textrm{k}$ (marked by point $\textrm{S}_\textrm{k}$ in the figure), the solver linearizes the IBR control equation around the solution of iteration $\#\textrm{k}$ based on
\begin{align}\label{eq:linearized_IBR}
\textrm{I}_\textrm{IBR}&=\textrm{I}_\textrm{IBR}[\textrm{k}]+f'(\textrm{V}_\textrm{IBR}[\textrm{k}-1])(\textrm{V}_\textrm{IBR}-\textrm{V}_\textrm{IBR}[\textrm{k}-1])\textrm{,}
\end{align}
where $f'(.)$ denotes the derivative of IBR control equation. Equation~(\ref{eq:linearized_IBR}) suggests that the linearized IBR model is a Norton equivalent in phasor domain given by
\begin{align}\label{eq:Norton}
\textrm{I}_\textrm{IBR}&=\textrm{I}_\textrm{n}[\textrm{k}]-\textrm{Y}_\textrm{n}[\textrm{k}]\cdot\textrm{V}_\textrm{IBR}\textrm{,}
\end{align}
whose Norton current phasor and admittance are given by
\begin{align}
\textrm{I}_\textrm{n}[\textrm{k}]&=\textrm{I}_\textrm{IBR}[\textrm{k}]+\textrm{Y}_\textrm{n}[\textrm{k}]\cdot\textrm{V}_\textrm{IBR}[\textrm{k}-1]\textrm{,}\label{eq:Norton_current}\\
\textrm{Y}_\textrm{n}[\textrm{k}]&=-f'(\textrm{V}_\textrm{IBR}[\textrm{k}-1])\textrm{,}\label{eq:Norton_admittance}
\end{align}
respectively. Figure~\ref{fig:Modified_IBR_Model} shows the proposed IBR model. Compared to the traditional model in Fig.~\ref{fig:Eq_Net_Trad_Sol}, the proposed model includes an additional shunt admittance in parallel with the ideal current source. The linearity of the proposed model facilitates the integration of IBR model equations with the linear equations of the network, enabling the solution of a concatenated linear system in each iteration.

The proposed solver assumes the availability of the derivative term $f'(\textrm{V}_\textrm{IBR})$ to calculate $\textrm{Y}_\textrm{n}[\textrm{k}]$ in (\ref{eq:Norton_current}) and (\ref{eq:Norton_admittance}). Depending on the available IBR data, two methods for calculating $\textrm{Y}_\textrm{n}[\textrm{k}]$ have been proposed, resulting in two variations of the proposed solver denoted as Solver 1 and Solver 2.

\subsection{Solver 1}
\label{section:Solver1}

This solver assumes the availability of a VCCS tabular IBR model that defines a piecewise linear relationship between $\textrm{I}_\textrm{IBR}$ and $\textrm{V}_\textrm{IBR}$. This relationship allows for the approximation of $\textrm{Y}_\textrm{n}[\textrm{k}]$ by the slope of the defined relationship. In iteration k, the amplitude of the voltage phasor in the previous iteration, $|\textrm{V}_\textrm{IBR}[\textrm{k}-1]|$, determines the selected segment of the piecewise linear characteristic. Assuming the amplitude lies between the voltages of the \textit{m}-\textit{th} and \textit{n}-\textit{th} rows of the table, denoted respectively by $V_m$ and $V_n$, one can write
\begin{align}\label{eq:Yn_VCCS}
\textrm{Y}_\textrm{n}[\textrm{k}]&=-\frac{I_m-I_n}{V_m-V_n}\textrm{,}~~~~~for~ V_m< |\textrm{V}_\textrm{IBR}[\textrm{k}-1]|<V_n\textrm{,}
\end{align}
where $I_m$ and $I_n$ denote the complex phasor of current for the \textit{m}-\textit{th} and \textit{n}-\textit{th} rows of the table, respectively. 

\begin{figure}[!t]
\centering
\includegraphics[width=0.36\textwidth, trim={0 0 0 0},clip]{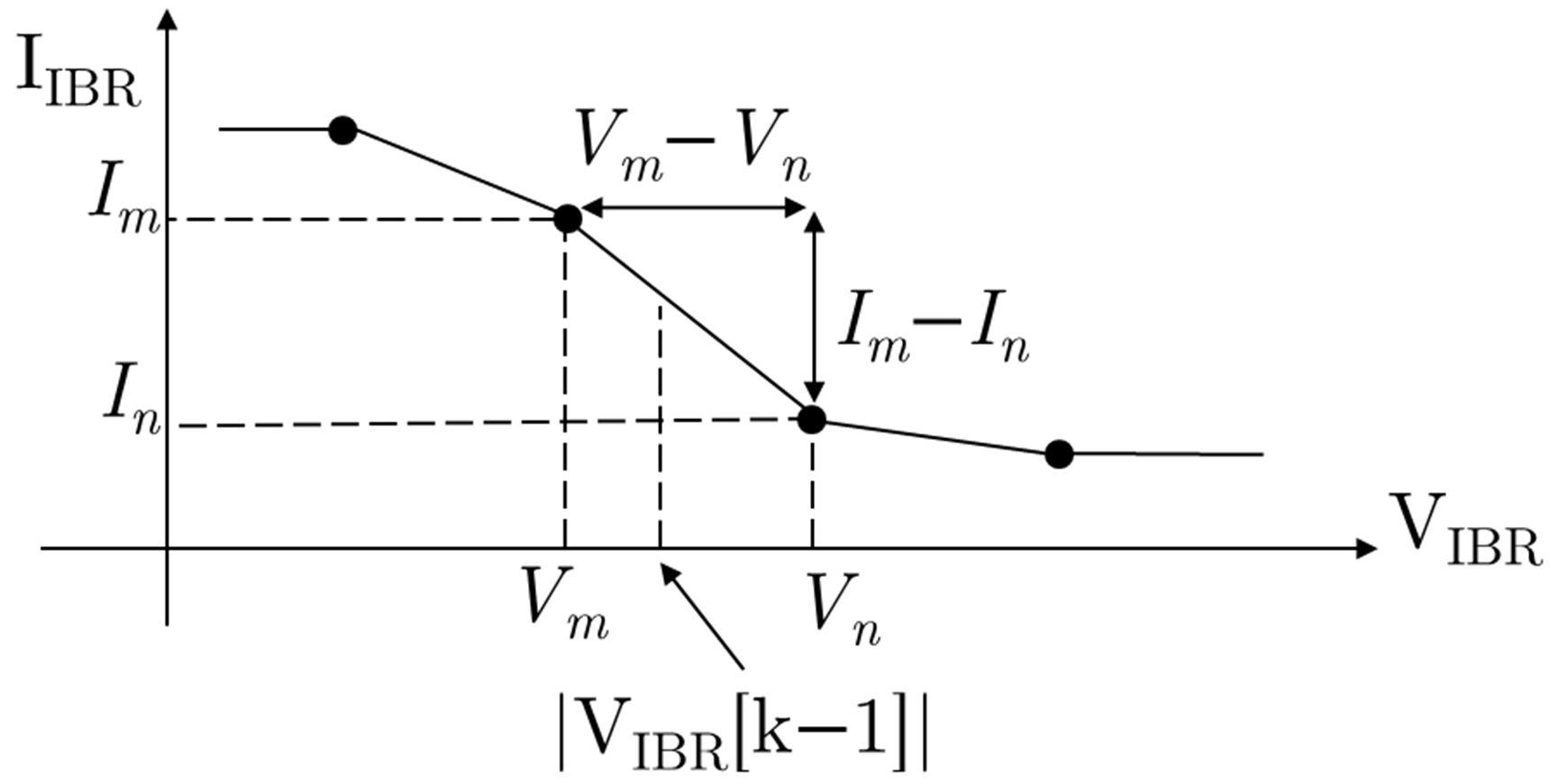}
\caption{Calculation of $\textrm{Y}_\textrm{n}[\textrm{k}]$ for an IBR VCCS tabular model.}
\label{fig:VCCS_Yn}
\end{figure}

\subsection{Solver 2}
\label{section:Solver2}

\begin{figure}[!t]
\centering
\includegraphics[width=0.38\textwidth, trim={0cm 0cm 0cm 0cm},clip]{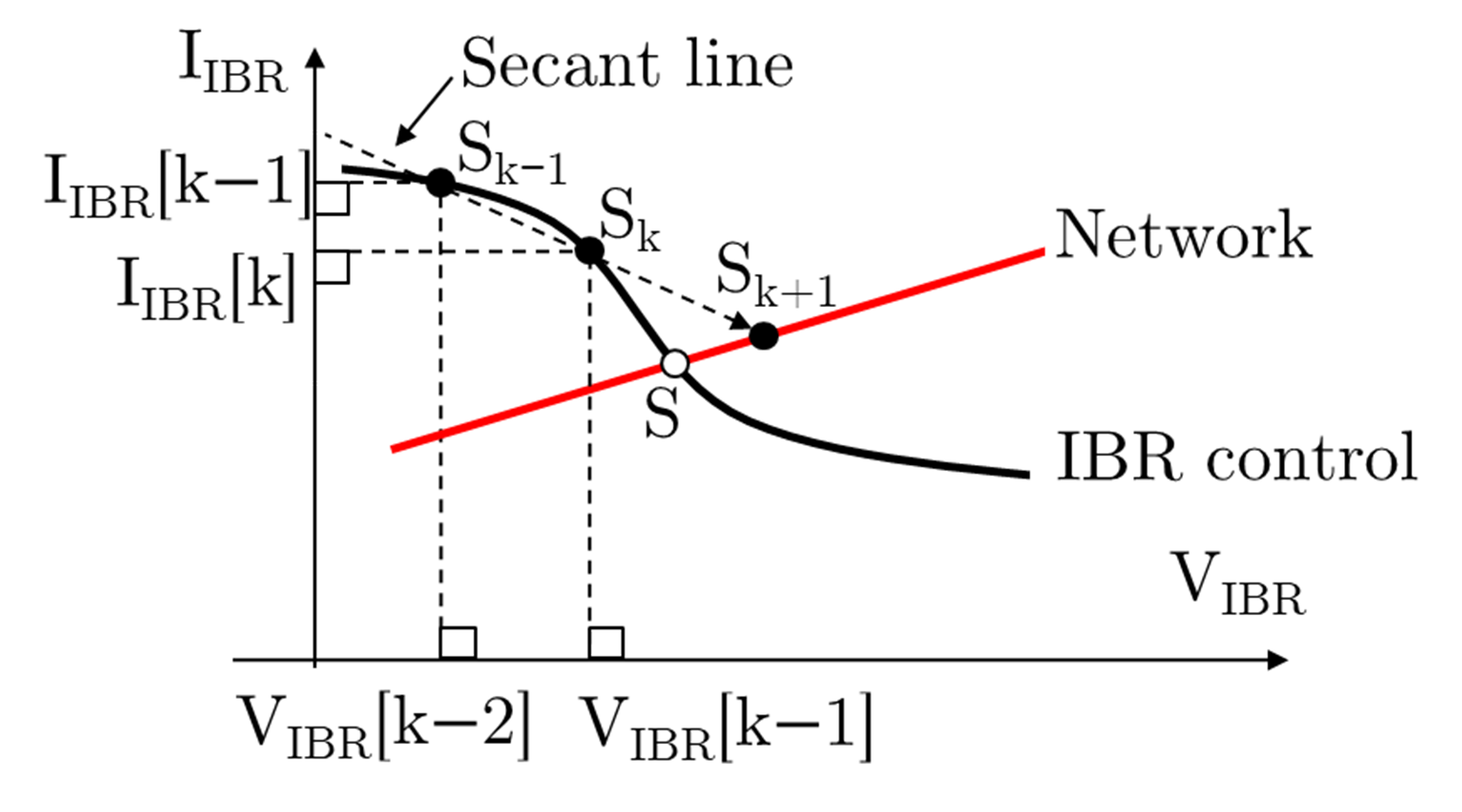}
\caption{Geometrical interpretation of the iteration of Solver 2.}
\label{fig:Yn_Secant}
\end{figure}

Solver 1 requires a piecewise linear relationship between $\textrm{I}_\textrm{IBR}$ and $\textrm{V}_\textrm{IBR}$ to be available to the solver. However, equation-based IBR models~\cite{Ref:C24} do not provide such a relationship. These equations are often proprietary and black-boxed, making them unavailable to the solver. To address this challenge, Solver 2 approximates $\textrm{Y}_\textrm{n}[\textrm{k}]$ based on the solutions of the preceding two iterations as follows
\begin{align}\label{eq:Yn_Secant}
\textrm{Y}_\textrm{n}[\textrm{k}]&=-\frac{\textrm{I}_\textrm{IBR}[\textrm{k}]-\textrm{I}_\textrm{IBR}[\textrm{k}-1]}{\textrm{V}_\textrm{IBR}[\textrm{k}-1]-\textrm{V}_\textrm{IBR}[\textrm{k}-2]}\textrm{.}
\end{align}
The geometrical interpretation involves replacing the tangent line with the secant line supported by $\textrm{V}_\textrm{IBR}[\textrm{k}-2]$ and $\textrm{V}_\textrm{IBR}[\textrm{k}-1]$, as shown in Fig.~\ref{fig:Yn_Secant}. This method requires two initial guesses, $\textrm{V}_\textrm{IBR}[0]$ and $\textrm{V}_\textrm{IBR}[1]$. At each step, only one evaluation of $\textrm{I}_\textrm{IBR}$ is necessary, because $\textrm{I}_\textrm{IBR}[\textrm{k}-1]$ is known from the previous iteration.

In contrast to Solver 1, Solver 2 is IBR-model agnostic meaning it does not require knowledge of the IBR control equation or the availability of IBR data in the VCCS tabular format. Therefore, it can be applied to a manufacturer-specific black-box model, a generic white-box equation-based IBR model, as well as a VCCS tabular model.

The appendix has studied the numerical convergence properties of Solver 1 and Solver 2, including both the mathematical proof and conditions necessary for convergence. Table~\ref{table:ComparisonConvSolv1Solv2} summarizes key properties. As shown, the order of convergence of Solver 1 and Solver 2 is 2 and approximately 1.62, respectively, indicating that Solver 1 demonstrates faster convergence. Furthermore, the convergence of both solvers depends on proper initialization; Equations (\ref{eq:ConvCondSolv1}) and (\ref{eq:ConvCondSolv2}) specify the initial conditions necessary for the convergence of Solver 1 and Solver 2, respectively.

\begin{table}[!t]\footnotesize
\renewcommand{\arraystretch}{1.3}
\caption{Summary of the Numerical Convergence Properties of Solver 1 and Solver 2.}
\label{table:ComparisonConvSolv1Solv2}
\centering
\begin{tabular}{m{10em}|m{8em}|m{8em}}
\hline
\textbf{Property} & \textbf{Solver 1} & \textbf{Solver 2} \\
\hline
Order of convergence & 2 & $\sim$1.62 \\
\hline
Initial condition necessary for convergence &	Equation~(\ref{eq:ConvCondSolv1}) &	Equation~(\ref{eq:ConvCondSolv2}) \\
\hline
\end{tabular}
\end{table}

\subsection{Solver Algorithm}
\label{section:Algorithm}

\begin{figure}[!t]
\centering
\includegraphics[width=0.48\textwidth, trim={0cm 0cm 0cm 0cm},clip]{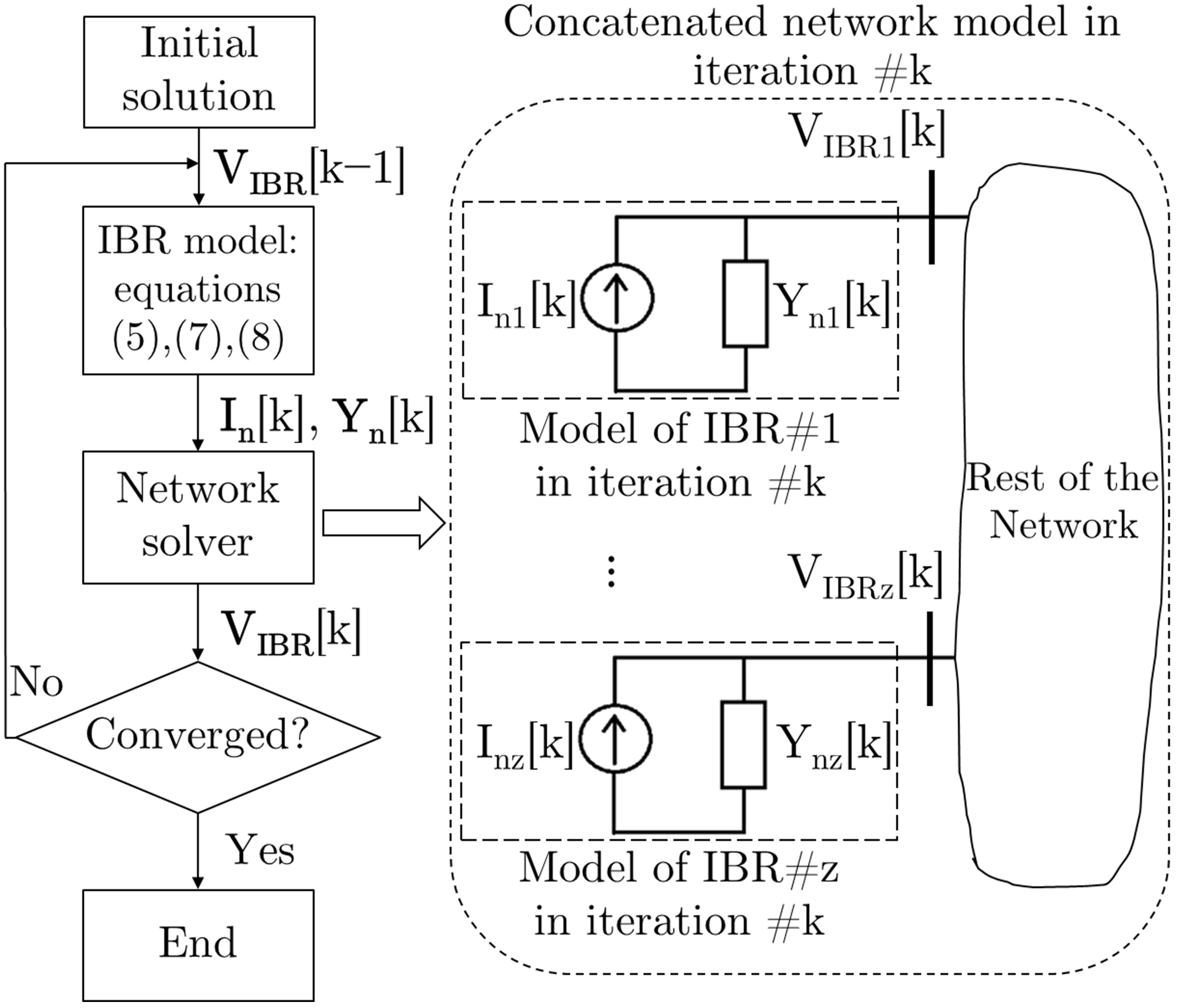}
\caption{The proposed solver algorithm.}
\label{fig:algorithm}
\end{figure}

Figure~\ref{fig:algorithm} shows a proposed algorithm that could be used for implementation of the proposed solvers in a fault analysis program. The following definitions have been used:
% \vspace{-2mm}
% \begin{align*}
%     z:\hspace{2mm}&\textrm{Total number of IBR models;} \\
%     \textrm{\textbf{V}}_\textrm{\textbf{IBR}}[\textrm{k}-1]:\hspace{2mm}&\{\textrm{V}_\textrm{IBR1}[\textrm{k}-1],...,\textrm{V}_\textrm{IBRz}[\textrm{k}-1]\}\textrm{, Vector of IBR}\\    
%     &\textrm{terminal voltage complex phasors in previous}\\
%     &\textrm{iteration;}\\
%     \textrm{\textbf{V}}_\textrm{\textbf{IBR}}[\textrm{k}]:\hspace{2mm}&\{\textrm{V}_\textrm{IBR1}[\textrm{k}],...,\textrm{V}_\textrm{IBRz}[\textrm{k}]\}\textrm{, Vector of IBR terminal}\\    
%     &\textrm{voltage complex phasors in the current}\\
%     &\textrm{iteration;}\\
%     \textrm{\textbf{I}}_\textrm{\textbf{n}}[\textrm{k}]:\hspace{2mm}&\{\textrm{I}_\textrm{n1}[\textrm{k}],...,\textrm{I}_\textrm{nz}[\textrm{k}]\}\textrm{, Vector of IBR Norton}\\    
%     &\textrm{current complex phasors in the current}\\
%     &\textrm{iteration;}\\
%     \textrm{\textbf{Y}}_\textrm{\textbf{n}}[\textrm{k}]:\hspace{2mm}&\{\textrm{Y}_\textrm{n1}[\textrm{k}],...,\textrm{Y}_\textrm{nz}[\textrm{k}]\}\textrm{, Vector of IBR Norton}\\    
%     &\textrm{admittance in the current iteration.}
% \end{align*}
\begin{align*}
    z=\hspace{1mm}&\textrm{Total number of IBR models;} \\
    \textrm{\textbf{V}}_\textrm{\textbf{IBR}}[\textrm{k}-1]=\hspace{1mm}&\{\textrm{V}_\textrm{IBR1}[\textrm{k}-1],...,\textrm{V}_\textrm{IBRz}[\textrm{k}-1]\}\textrm{;}\\
    \textrm{\textbf{V}}_\textrm{\textbf{IBR}}[\textrm{k}]=\hspace{1mm}&\{\textrm{V}_\textrm{IBR1}[\textrm{k}],...,\textrm{V}_\textrm{IBRz}[\textrm{k}]\}\textrm{;}\\
    \textrm{\textbf{I}}_\textrm{\textbf{n}}[\textrm{k}]=\hspace{1mm}&\{\textrm{I}_\textrm{n1}[\textrm{k}],...,\textrm{I}_\textrm{nz}[\textrm{k}]\}\textrm{; and}\\
    \textrm{\textbf{Y}}_\textrm{\textbf{n}}[\textrm{k}]=\hspace{1mm}&\{\textrm{Y}_\textrm{n1}[\textrm{k}],...,\textrm{Y}_\textrm{nz}[\textrm{k}]\}\textrm{.}
\end{align*}
The algorithm begins with an initial guess for IBR terminal voltages. During iteration $\#\textrm{k}$, the voltages $\textrm{\textbf{V}}_\textrm{\textbf{IBR}}[\textrm{k}-1]$ are processed by individual IBRs. Subsequently, $\textrm{\textbf{Y}}_\textrm{\textbf{n}}[\textrm{k}]$ is computed using either (\ref{eq:Yn_VCCS}) for Solver 1 or (\ref{eq:Yn_Secant}) for Solver 2. Following this, $\textrm{\textbf{I}}_\textrm{\textbf{n}}[\textrm{k}]$ is determined from~(\ref{eq:Norton_current}). The IBR models are then updated and integrated with the rest of the network to form a concatenated network model. The voltages $\textrm{\textbf{V}}_\textrm{\textbf{IBR}}[\textrm{k}]$ are derived from the solution of this concatenated network. Finally, convergence is checked, and if the stopping criteria are not met, the steps are repeated.

\section{Case Studies and Simulation Results}

The performance of Solver 1 and Solver 2 has been evaluated against that of the traditional SC solver of Section~\ref{sec:Trad_solver} on the test system of Fig.~\ref{fig:simpleTestSystem} and a multi-IBR test system depicted in Fig.~\ref{fig:MultiIBR}. The stopping criteria for the iterative method of the solvers have been defined as either a 5\% tolerance on the change in the amplitude of $\textrm{V}_\textrm{IBR}$ with respect to the previous iteration, or a maximum of 20 iterations. Two initialization methods have been considered: ``initialization from zero'' which calculates $\textrm{\textbf{V}}_\textrm{\textbf{IBR}}[\textrm{0}]$ by ignoring IBR current injection ($\textrm{\textbf{I}}_\textrm{\textbf{IBR}}[\textrm{0}]=\textbf{0}$), and ``initialization from pre-fault power flow'' which uses the pre-fault power flow solution to determine $\textrm{\textbf{V}}_\textrm{\textbf{IBR}}[\textrm{0}]$. For Solver 1 and Solver 2, the considered initial conditions have been checked against the convergence criteria specified in the appendix to ensure convergence. The phasor-domain simulation results of the solvers have been cross-examined against EMT simulation results. In the figures to follow, these EMT results have been marked by a dashed black line.

\subsection{Case 1: Single-IBR Test System}
\label{sec:CaseStudy1}

\begin{figure}[!t]
\centering
\includegraphics[width=0.53\textwidth, trim={0.25cm 0cm 0cm 0cm},clip]{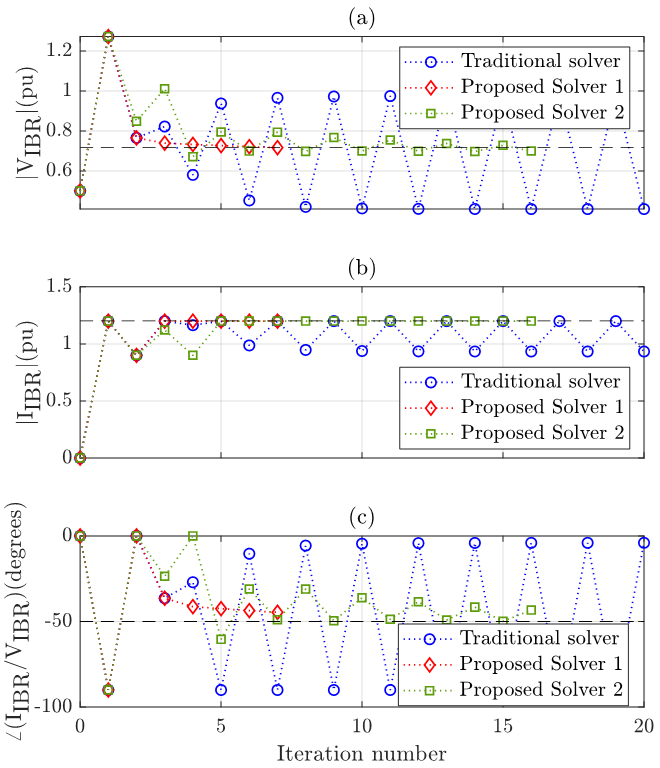}
\caption{Comparison of the iteration of the traditional solver, Solver 1, and Solver 2 with initialization from zero on the test system of Fig.~\ref{fig:simpleTestSystem}: (a) amplitude of the complex phasor of IBR positive-sequence terminal voltage; (b) amplitude of the complex phasor of IBR positive-sequence output current; and (c) phase angle of the complex phasor of IBR positive-sequence output current relative to the complex phasor of IBR positive-sequence terminal voltage. The dashed black line represents the EMT solution.}
\label{fig:comparison_solvers_simpletestsyst}
\end{figure}

Fault $\textrm{F}_1$ in the test system of Fig.~\ref{fig:simpleTestSystem} has been repeated. Figure~\ref{fig:comparison_solvers_simpletestsyst} illustrates the  iterations of the traditional solver, Solver 1, and Solver 2, assuming initialization from zero. As shown, the traditional solver does not converge; however, Solver 1 and Solver 2 converge in 7 and 16 iterations, respectively, matching the EMT solution. The results suggest the improved numerical stability of the proposed solver. The faster convergence of Solver 1 compared to Solver 2 is consistent with its higher order of convergence, as presented in Table~\ref{table:ComparisonConvSolv1Solv2}. 

The plot of $\textrm{V}_\textrm{IBR}$ in Fig.~\ref{fig:comparison_solvers_simpletestsyst} shows an overshoot of about 1.3 pu in iteration 1. The reason is the abrupt change of $\textrm{I}_\textrm{IBR}$ from zero in iteration 0 to a purely reactive current of 1.20$\angle-\textrm{90}{\degree}$pu in iteration 1. This abrupt change of current leads to a voltage overshoot due to a weak network condition at IBR terminal. This artificial overshoot is not numerically desirable since it may make convergence more challenging. Further, the high voltage may be outside the validity range of the FRT control programmed in the IBR model, thus leading to an incorrect solution. In tests, this overshoot was eliminated by initializing from pre-fault power flow. 

Table~\ref{table:Comp_iterations_Solvers1} presents the number of iterations of the solvers under the two initialization methods. Both Solver 1 and Solver 2 achieve convergence in all scenarios, with Solver 1 demonstrating faster convergence. Initialization from pre-fault power flow reduced the number of iterations of Solver 2 from 16 to 10; however, it increased the number of iterations of Solver 1 from 7 to 9. The result suggests that while initialization from pre-fault power flow may reduce potential numerical oscillations, it may not improve the speed of convergence.

\begin{figure}[!t]
\centering
\includegraphics[width=0.5\textwidth]{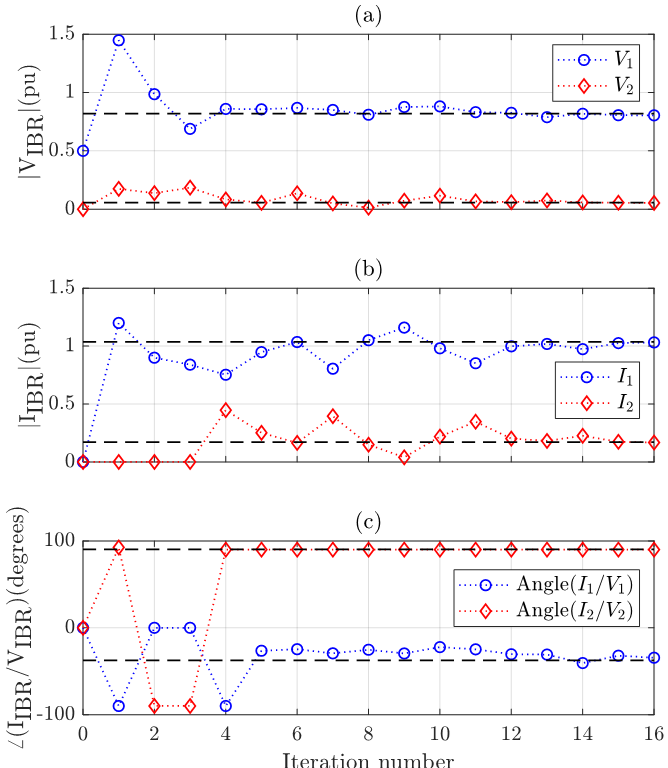}
\caption{The iteration of Solver 2 for an unbalanced fault in the test system of Fig.~\ref{fig:simpleTestSystem}: (a) amplitude of the complex phasor of IBR positive- and negative-sequence terminal voltage ($V_1$ and $V_2$); (b) amplitude of the complex phasor of IBR positive- and negative-sequence output current ($I_1$ and $I_2$); and (c) phase angle of the complex phasor of IBR positive- and negative-sequence output current relative to the complex phasor of IBR positive- and negative-sequence terminal voltage. The dashed black line represents the EMT solution.}
\label{fig:unbalanced_simple}
\end{figure}

\begin{table}[!t]\footnotesize
\renewcommand{\arraystretch}{1.3}
\caption{Comparison of the Number of Iterations of the Traditional Solver, Solver 1, and Solver 2 in the Test System of Fig.~\ref{fig:simpleTestSystem}.}
\label{table:Comp_iterations_Solvers1}
\centering
\begin{tabular}{m{8em}|m{9em}|m{9em}}
\hline
\multirow{2.75}{*}{\textbf{Solver}} & \multicolumn{2}{c}{\textbf{Number of Iterations}}\\ \cline{2-3}
 & \textbf{Initialization from zero} & \textbf{Initialization from pre-fault power flow} \\
\hline
Traditional solver & No convergence & No convergence\\
\hline
Solver 1 & 7 & 9 \\
\hline
Solver 2 & 16 & 10\\
\hline
\end{tabular}
\end{table}

The proposed solvers are applicable to unbalanced faults. To illustrate that, a phase-A-to-B-to-ground fault in the test system of Fig.~\ref{fig:simpleTestSystem} has been solved using Solver 2. The IBR has been represented by an equation-based model~\cite{Ref:Type4Model} with negative-sequence current control based on a k-factor control~\cite{Ref:WesPES23} conforming with IEEE Std. 2800-2022~\cite{Ref:2800}. Figure~\ref{fig:unbalanced_simple} presents the iteration of Solver 2 for this unbalanced fault. The solver converges in 16 iterations, matching the EMT solution. The solution is consistent with the IBR control; for a negative-sequence terminal voltage of $V_2=\textrm{0.06}$ pu, the IBR injects a negative-sequence current of $I_2=\textrm{0.17}\angle\textrm{90}\degree$ pu, conforming with the relevant requirements of IEEE Std. 2800-2022.

\subsection{Case 2: Multi-IBR Test System}
\label{sec:CaseStudy2}

\begin{figure}[!t]
\centering
\includegraphics[width=0.48\textwidth, trim={0cm 0cm 0cm 0cm},clip]{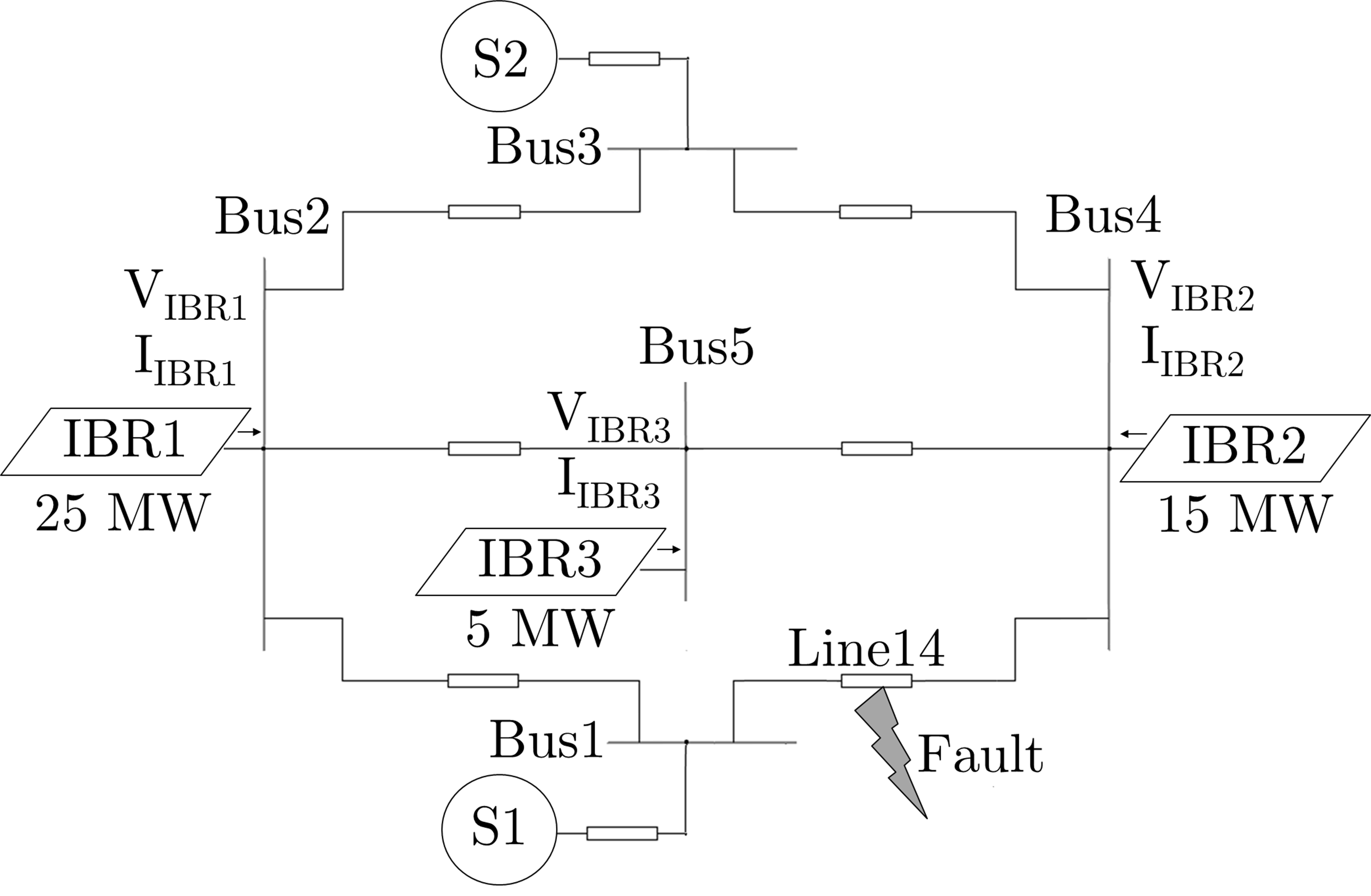}
\caption{The multi-IBR test system of Section~\ref{sec:CaseStudy2}.}
\label{fig:MultiIBR}
\end{figure}

To demonstrate its applicability to a multi-IBR power system, the proposed solver has been applied to the test system shown in Fig.~\ref{fig:MultiIBR}, representing a portion of a transmission network. Three solar PV-based parks, labeled IBR1, IBR2, and IBR3, with installed capacities of \{25, 15, 5\} MW, respectively, have been connected to buses Bus2, Bus4, and Bus5, respectively. The terminal voltages and currents of the IBRs have been denoted by $\textrm{V}_\textrm{IBRi}$ and $\textrm{I}_\textrm{IBRi}$, where the subscript ``i'' indicates the IBR index ($\textrm{i}$=\{1, 2, 3\}). The rest of the transmission system has been represented by two voltage sources, S1 and S2, each behind their respective impedances. A bolted three-phase-to-ground fault has been applied in the middle of line Line14. IBR terminal voltages and currents have been calculated using the traditional solver, Solver 1, and Solver 2.

\begin{figure}[!t]
\centering
\includegraphics[width=0.5\textwidth]{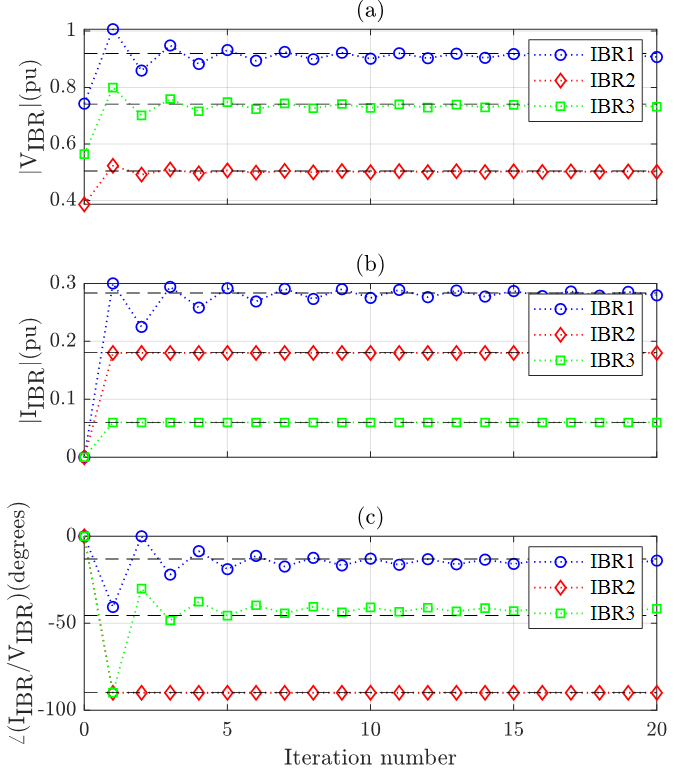}
\caption{Iterations of the traditional solver with initialization from zero on the test system of Fig.~\ref{fig:MultiIBR}: (a) amplitude of the complex phasor of IBR positive-sequence terminal voltage; (b) amplitude of the complex phasor of IBR positive-sequence output current; and (c) phase angle of the complex phasor of IBR positive-sequence output current relative to the complex phasor of IBR positive-sequence terminal voltage. The dashed black line represents the EMT solution.}
\label{fig:MultiIBR_TradSolver}
\end{figure}

\begin{figure}[!t]
\centering
\includegraphics[width=0.5\textwidth]{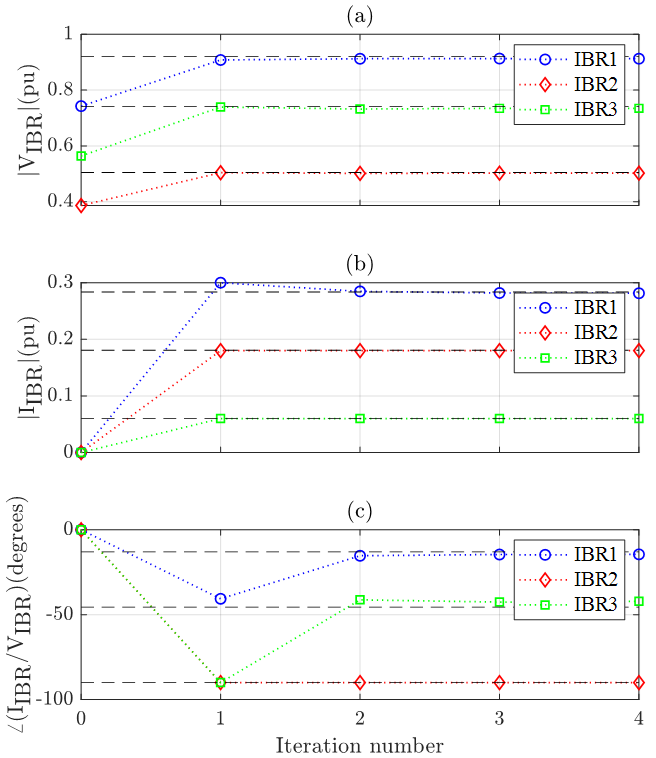}
\caption{The iteration of Solver 1 with initialization from zero on the test system of Fig.~\ref{fig:MultiIBR}: (a) amplitude of the complex phasor of IBR positive-sequence terminal voltage; (b) amplitude of the complex phasor of IBR positive-sequence output current; and (c) phase angle of the complex phasor of IBR positive-sequence output current relative to the complex phasor of IBR positive-sequence terminal voltage. The dashed black line represents the EMT solution.}
\label{fig:MultiIBR_Solver1}
\end{figure}

\begin{figure}[!t]
\centering
\includegraphics[width=0.5\textwidth]{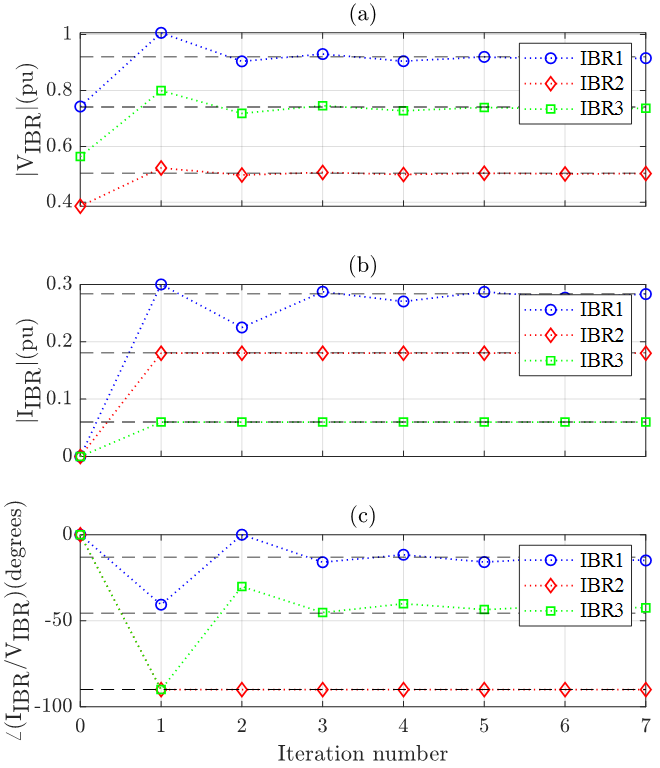}
\caption{The iteration of Solver 2 with initialization from zero on the test system of Fig.~\ref{fig:MultiIBR}: (a) amplitude of the complex phasor of IBR positive-sequence terminal voltage; (b) amplitude of the complex phasor of IBR positive-sequence output current; and (c) phase angle of the complex phasor of IBR positive-sequence output current relative to the complex phasor of IBR positive-sequence terminal voltage. The dashed black line represents the EMT solution.}
\label{fig:MultiIBR_Solver2}
\end{figure}

Figures~\ref{fig:MultiIBR_TradSolver}, ~\ref{fig:MultiIBR_Solver1}, ~\ref{fig:MultiIBR_Solver2} show the iteration of the traditional solver, Solver 1, and Solver 2, respectively, assuming initialization from zero. As shown, the traditional solver takes 20 iterations to converge, whereas Solver 1 and Solver 2 converge in 4 and 7 iterations, matching the EMT solution. The results suggest the improved convergence of the proposed solvers, with Solver 1 showing superior convergence speed compared to Solver 2, as expected. 

Table~\ref{table:Comp_iterations_Solvers2} tabulates the number of iterations of the solvers under the two initialization methods. As shown, Solver 1 and Solver 2 exhibit faster convergence than the traditional solver, with Solver 1 being the fastest. Initialization from power flow improves the convergence speed of the traditional solver and Solver 2; however, it does not improve the convergence speed of Solver 1. The results support the conclusions of Section~\ref{sec:CaseStudy1} regarding the improved convergence of Solver 1 and Solver 2 compared to the traditional solver and the impact of initialization on convergence speed.

\begin{table}[!t]\footnotesize
\renewcommand{\arraystretch}{1.3}
\caption{Comparison of the Number of Iterations of the Traditional Solver, Solver 1, and Solver 2 in the Test System of Fig.~\ref{fig:MultiIBR}}
\label{table:Comp_iterations_Solvers2}
\centering
\begin{tabular}{m{8em}|m{9em}|m{9em}}
\hline
\multirow{2.75}{*}{\textbf{Solver}} & \multicolumn{2}{c}{\textbf{Number of Iterations}}\\ \cline{2-3}
 & \textbf{Initialization from zero} & \textbf{Initialization from pre-fault power flow} \\
\hline
Traditional solver & 20 & 19 \\
\hline
Solver 1 & 4 & 4 \\
\hline
Solver 2 & 7 & 4\\
\hline
\end{tabular}
\end{table}

\section{Conclusion}
A Newton-Raphson (NR)-based numerical solver has been proposed to integrate Inverter-Based Resource (IBR) short circuit (SC) models into phasor-domain fault analysis programs. Two variations of the proposed solver, denoted as Solver 1 and Solver 2, have been developed, each tailored to different types of IBR SC models. Solver 1, which assumes the availability of a voltage-controlled current source (VCCS) tabular model, implements the iterative NR method. In contrast, Solver 2 is model-agnostic and employs a secant-based variation of the NR method. An algorithm for implementing these solvers in phasor-domain fault analysis programs has been proposed. Simulations have demonstrated that both solvers exhibit superior numerical stability and convergence properties compared to a traditional SC solver. While Solver 1 shows faster convergence, its application is limited to the VCCS tabular IBR model. Solver 2, although slower in convergence, can handle both VCCS tabular and equation-based IBR models. Simulations indicate that initialization is crucial for achieving convergence, though it may not necessarily enhance the speed of convergence. Case studies have further demonstrated the effectiveness of these solvers in simulating a multi-IBR power system, including scenarios with unbalanced faults.

\appendix[Convergence Analysis]
\label{sec:appendix}

The convergence of the proposed solver can be established by employing the general convergence theorem delineated in~\cite{Ref:BookStoer}. 

\noindent\textbf{Theorem}: Let $\Phi$ be an iteration function on $\mathbb{C}^n$ defined by $x_{p+1}=\Phi(x_p)$ for $p=\{0,1,2,...\}$. Let $\xi$ be a fixed point of $\Phi$. For all initial vectors $x_0$ taken from a neighborhood $N(\xi)$ and for the generated sequence of $\Phi$, let an inequality of the form ${\Vert}x_{p+1}–\xi{\Vert}\leq C{\Vert}x_p–\xi{\Vert}^\kappa$ hold for all $p\geq0$, where $C<1$ if $\kappa=1$, and ${\Vert}.{\Vert}$ is a norm measuring the distance between two vectors on $\mathbb{C}^n$. The iteration method defined by $\Phi$ is said to be a method of at least $\kappa$\textit{-th} order for determining $\xi$. It can be shown that each method of at least $\kappa$\textit{-th} order for determining a fixed point $\xi$ is locally convergent, in the sense that there is a neighborhood $N(\xi)$ of $\xi$ with the property that for all initial $x_0\in N(\xi)$, the sequence generated by $\Phi$ converges to $\xi$.

The convergence of Solver 1 and Solver 2 can be demonstrated as specific cases of the general convergence theorem on $\mathbb{C}$. Let $g(.)$ be a nonlinear function defined as
\begin{align}\label{eq:gx}
g(x)=f(x)-\frac{x}{\textrm{Z}_\textrm{th}}+\frac{\textrm{V}_\textrm{th}}{\textrm{Z}_\textrm{th}}
\textrm{,}
\end{align}
where $f(.)$ is the nonlinear IBR control equation defined in~(\ref{eq:IBR_control_eq}).

\subsection{Convergence of Solver 1}
The iteration function of Solver 1 is defined by 
\begin{align}\label{eq:Phi_Solver1}
\Phi(x_p)=x_p–\frac{g(x_p)}{g'(x_p)}\textrm{,}
\end{align}
with $g(.)$ given in (\ref{eq:gx}). Let $\xi$ be a simple root of the nonlinear equation $g(x)=0$. It can be shown that there exists a $r>0$ such that for every $x_0\in [\xi-r, \xi+r]$, the iterative sequence $\{x_0, x_1, …, x_p, …\}$ quadratically converges to $\xi$ (i.e., $\kappa=\textrm{2}$), subject to the following conditions~\cite{Ref:BookStoer}:

\noindent\textit{Condition 1}: $g(x)$ has a zero at $\xi$. For this condition to be satisfied, (\ref{eq:IBR_control_eq}) and (\ref{eq:network_KVL}) must allow a solution. If no solution exists, then the iteration does not converge. Reference~\cite{Ref:C24} has presented an example of such a numerical non-convergence for a bolted three-phase fault at IBR terminal for which the power factor of IBR current injection is inconsistent with that of the fault loop impedance. Reference~\cite{Ref:ManishNumerical} has presented another example wherein an IBR control-driven instability of the actual power system results in numerical non-convergence of the solver. In such cases, the non-convergence is the correct solution since no stable operating point exists in the actual power system;

\noindent\textit{Condition 2}: $g'(x_0)\neq0$. For this condition to be satisfied, the initial guess needs to be selected such that $g'(\textrm{V}_\textrm{IBR}[0])\neq0$ which requires that 
\begin{align}\label{eq:ConvCondSolv1}
        f'(\textrm{V}_\textrm{IBR}[0])\neq\frac{1}{\textrm{Z}_\textrm{th}}\textrm{.}
\end{align}
Geometrically, this implies that in Fig.~\ref{fig:Geometric_NR}, the tangent line at the point $\textrm{V}_\textrm{IBR}[0]$ should not be parallel to the network curve. Violation of this condition means that the first tangent line does not intersect the network curve, and the iteration cannot be set up.
    
\begin{figure}[!t]
\centering
\includegraphics[width=0.5\textwidth]{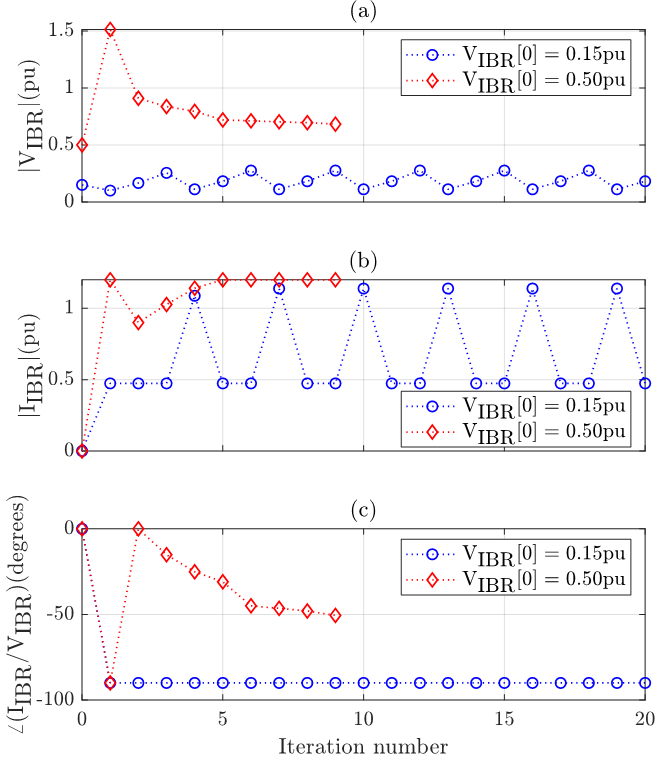}
\caption{Impact of initialization on the convergence of Solver 1 showing non-convergence under $\textrm{V}_\textrm{IBR}[0]=\textrm{0.15}$pu and convergence under $\textrm{V}_\textrm{IBR}[0]=\textrm{0.50}$pu: (a) amplitude of the complex phasor of IBR positive-sequence terminal voltage; (b) amplitude of the complex phasor of IBR positive-sequence output current; and (c) phase angle of the complex phasor of IBR positive-sequence output current relative to the complex phasor of IBR positive-sequence terminal voltage.}
\label{fig:NumericalNonConvCond2}
\end{figure}

To illustrate an example of non-convergence due to the violation of these conditions, the test system of Fig~\ref{fig:simpleTestSystem} has been considered. The IBR model is derived from Table~\ref{table:VCCS}, with the current amplitudes in the last two rows adjusted to 0.9479 pu (second-to-last row) and 0.4739 pu (last row). The IBR tie-line impedance has been changed to j0.2 pu, and fault $\textrm{F}_\textrm{1}$ has been simulated using Solver 1 with two initial solutions of $\textrm{V}_\textrm{IBR}[0]=\{\textrm{0.15}, \textrm{0.50}\}$ pu; it can be shown that the first value violates Condition 2 whereas the second value satisfies it. Figure~\ref{fig:NumericalNonConvCond2} presents the results, suggesting non-convergence for $\textrm{V}_\textrm{IBR}[0]=\textrm{0.15}$ pu and convergence for $\textrm{V}_\textrm{IBR}[0]=\textrm{0.50}$ pu, as expected. The case study suggests that proper initialization is crucial for the convergence of the proposed solver.

\subsection{Convergence of Solver 2}

The iteration function of Solver 2 is defined by 
\begin{align}\label{eq:Phi_Solver2}
\Phi(x_p)=x_p–g(x_p)\cdot\frac{x_p–x_{p–1}}{g(x_p)–g(x_{p–1})}\textrm{,}
\end{align}
with $g(x)$ given in (\ref{eq:gx}). Let $\xi$ be a simple root of the nonlinear equation $g(x)=0$. It can be proven that there exists a $r>0$ such that for every $x_0\in [\xi-r, \xi+r]$, the iterative sequence $\{x_0, x_1, …, x_p, …\}$ converges to $\xi$ with an order of convergence of $\kappa=\frac{(\sqrt{5}+1)}{2}\approx1.62$. The conditions for convergence are:

\noindent\textit{Condition 1}: $g(x)$ has a root at $\xi$. This condition shares the same implications as those of \textit{Condition 1} of Solver 1;

\noindent\textit{Condition 2}: $g(x_0)\neq g(x_1)$. This condition requires that the initial solutions, denoted as $\textrm{V}_\textrm{IBR}[0]$ and $\textrm{V}_\textrm{IBR}[1]$ must be distinct from one another, i.e.,
\begin{align}\label{eq:ConvCondSolv2}
    \textrm{V}_\textrm{IBR}[0]\neq\textrm{V}_\textrm{IBR}[1]\textrm{.}
\end{align}

\end{document}